\documentclass{revtex4}
\usepackage{amssymb}
\usepackage{amssymb,amsmath}
 \usepackage{graphicx,epsfig}

\begin{document}

\title{A New Constraint for the Coupling of Axion-like particles to Matter
via an Ultra-Cold Neutron Gravitational Experiments}
\author{S. Bae{\ss}ler}
 \affiliation {Institut of physics,
University of Mainz, 55099 Mainz, Germany}
\author{V.V. Nesvizhevsky}
\affiliation{Institut Laue-Langevin (ILL), 6 rue Jules Horowitz,
 F-38042, Grenoble, France}
\author{K.V. Protasov}
 \affiliation {Laboratoire de Physique Subatomique et de Cosmologie
 (LPSC), IN2P3-CNRS, UJFG, 53, Avenue des Martyrs, F-38026, Grenoble,
 France}
\author{A.Yu. Voronin}
\affiliation{P.N. Lebedev Physical Institute, 53, Leninsky
prospect, 117924, Moscow, Russia}

\begin{abstract}
We present a new constraint for the axion monopole-dipole coupling
in the range of 1 $\mu$m -- a few mm, previously unavailable for
experimental study. The constraint was obtained using our recent
results on the observation of neutron quantum states in the
Earth's gravitational field. We exploit the ultimate sensitivity
of ultra-cold neutrons (UCN) in the lowest gravitational states
above a material surface to any additional interaction between the
UCN and the matter, if the characteristic interaction range is
within the mentioned domain. In particular, we find that the upper
limit for the axion monopole-dipole coupling constant is
$g_pg_s/(\hbar c)<2$ $10^{-15}$ for the axion mass in the
``promising'' axion mass region $M_A\sim 1$ meV.

\end{abstract}

\maketitle

A vanishing value for the neutron electric dipole moment motivated
the introduction of hypothetical light (pseudo) scalar bosons
(commonly called axions), as an extension of the Standard Model
\cite{AxTheo1, AxTheo2,AxTheo3,AxTheoRev}. According to the
suggested theories the axion mass could be in the range of
$10^{-6}<M_A<10^{-1}$ eV, while its coupling to photons, leptons
and nucleons is not fixed by the existing models (though it is
extremely weak). Following the theoretical predictions mentioned
intensive searches for axions have been performed over recent
decades. These studies include testing the astrophysical
consequences of the axion theories, QED effects (axion-two photon
coupling) and macroscopic forces (spin-matter coupling). They put
severe constraints on axion-matter coupling in different axion
mass ranges. A detailed review of axion studies can be found in
\cite{Rosenberg, PDG}.

The recently reported positive results of the PVLAS experiment on
light polarization rotation in a vacuum in the presence of a
transverse magnetic field \cite{PVLAS} may be seen as evidence of
the long sought axion \cite{NatureLam}. According to \cite{PVLAS},
the mass of neutral boson possibly responsible for the observed
signal is $1<M_A<1.5$ meV.

The value of the axion-photon coupling strength obtained from the
PVLAS experiment is in contradiction with recent CAST observations
\cite{CAST}. Several ideas have been discussed recently, in
\cite{Anticontr,Antoniadis}, capable of explaining this
discrepancy.

This intriguing result makes it particularly important to carry
out independent testing on the axion-matter coupling in the
corresponding distance range of $130< \lambda < 200~\mu$m.

In the present Letter we report on constraints for axion
monopole-dipole coupling. Such coupling results in a spin-matter
CP violating Yukawa-type interaction potential \cite{Moody}
\begin{eqnarray}
V(\vec{r})= \hbar g_p g_s \frac{\vec{\sigma}\cdot \vec{n}}{8\pi
mc} \left(\frac{1}{\lambda r}+ \frac{1}{r^2} \right)
\mbox{e}^{-r/\lambda}\label{potelem}
\end{eqnarray}
between spin and matter, where $g_p g_s$  is the product of
couplings at the scalar and polarized vertices and $\lambda$ is
the force range. Here $r $ is the distance between a neutron and a
nucleus, $\vec{n}=\vec{r}/r$ is a unitary vector, and $m$ is the
nucleon mass.

Only a few experiments for distances below 100 mm have placed
upper limits on the product coupling in a system of magnetized
media and test masses \cite{PDG}. One experiment \cite{Yuod} had
peak sensitivity at $\sim 100$ mm and two other ones \cite{Ritter,
Ni} had peak sensitivity at $\sim 10$ mm.

The constraint for the $g_sg_p$ presented in this article is
competitive in the distance range of $1<\lambda<10^3$ $\mu m$.

In the experimental method used \cite{Nature1,PRD1,EPJC,NIM} UCN
move above a nearly perfect horizontal mirror in the presence of
the Earth's gravitational field. A combination of a mirror and the
gravitational potential binds neutrons close to the mirror surface
in the so-called gravitational states. The characteristic scale of
this problem is $l_0=\sqrt[3]{ \hbar^2/(2m^{2}g)}=5.87$ $\mu m$,
while the characteristic size of the lowest gravitational neutron
state (a quasi classical turning point height) is  $\sim 2.4 l_0
=13.7~\mu$m. The neutron spatial distribution, directly measured
in our experiment, turns out to be very sensitive to any
additional potential, with a characteristic range from fractions
to tens of $l_0$. This property of the neutron states enables us
to establish a new limit on $g_sg_p$. In the following we will
describe the experimental setup and the estimation of the
axion-mediated interaction intensity constraint.

\begin{figure}
  \centering
 \includegraphics[width=115mm]{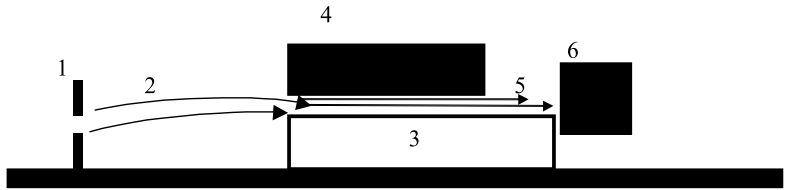}
\caption{The general experimental scheme. From the left to right:
vertical solid lines indicate two plates of the entrance
collimator (1); solid arrows show classical neutron trajectories
(2) between the collimators and the entrance to the slit between a
mirror (3, grey rectangle on bottom) and a scatterer (4, black
rectangle on top); Dashed horizontal lines show quantum motion of
neutrons above the mirror (5); black box indicates a neutron
detector (6). The size of the slit between the mirror and the
scatterer can be finely tuned and measured.}\label{FigInstall}
\end{figure}

The experiment \cite{PRD1,EPJC,NIM} involved the measurement of
the neutron flux through the horizontal gap (slit) between a
horizontal mirror (below) and a scatterer (above), as a function
of the slit size $\Delta h$ (see figure~\ref{FigInstall}).

\begin{figure}
  \centering
 \includegraphics[width=115mm]{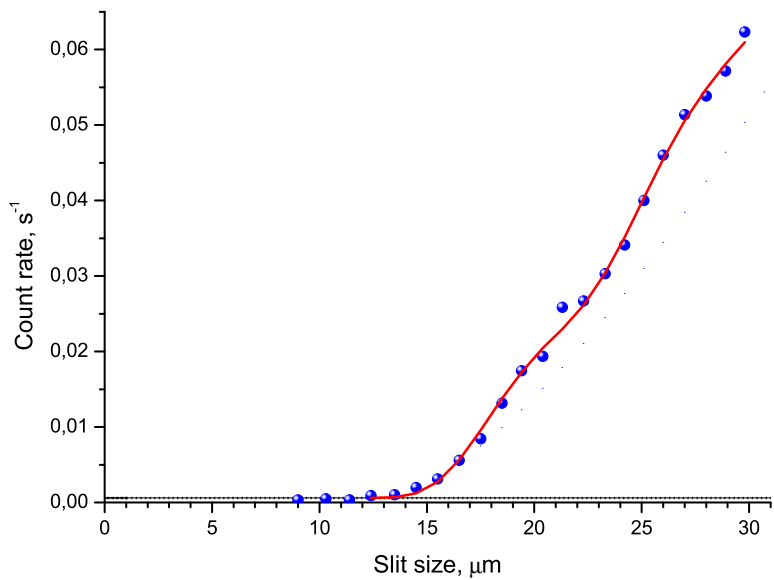}
\caption{A dependence of the neutron flux through a slit between
the mirror and the scatterer versus the slit size. The circles
show the data points, the curve is the theoretical description
within the quasi classical approach. The horizontal lines indicate
the detector background and its uncertainty.}\label{QuantumMotion}
\end{figure}

The aim of the experiment was to demonstrate, for the first time,
the existence of the quantum states of matter in a gravitational
field. An example of the dependence of the neutron flux on the
slit size $\Delta h$ is presented in figure~\ref{QuantumMotion}
\cite{EPJC}.

This dependence is sensitive to the presence of quantum states of
neutrons in the potential well formed by the Earth's gravitational
field and the mirror. In particular, the neutron flux was found to
be equal to zero within the experimental accuracy if the slit size
$\Delta h$ is smaller than the characteristic spatial size (a
quasi classical turning point height) of the lowest quantum state
of $\sim 15 \mu$m in this potential well.

This flux was fitted by a quasi classical function \cite{EPJC} and
the results for the two lowest quantum states
\begin{eqnarray}
&&z_1^{\mbox{\tiny exp}}= 12,2 \pm 0,7_{\mbox{\tiny stat}} \pm
1,8_{\mbox{\tiny sys}} \mu{\mbox{m}}, \nonumber \\
&&z_2^{\mbox{\tiny exp}}= 21,6 \pm 0,7_{\mbox{\tiny stat}}\pm
2,2_{\mbox{\tiny sys}} \mu{\mbox{m}},\label{zexp}
\end{eqnarray}
are in agreement (25\%) with the expected values
\begin{eqnarray}
&&z_1^{\mbox{\tiny qc}}= 13,7\mu{\mbox{m}}, \nonumber\\
&&z_2^{\mbox{\tiny qc}}= 24,0 \mu{\mbox{m}}.\label{zth}
\end{eqnarray}

It should also be mentioned that the method  used in this
experiment (based on position sensitive detectors) to visualize
the wave functions of the low lying states \cite{EPJC,PANIC} also
revealed no deviation from expected theoretical behavior.


An additional interaction (\ref{potelem}) between a neutron and a
mirror's nuclei produces an additional neutron-mirror interaction
potential.

If the mirror's density is constant and equal to $\rho_m$, an
additional potential of the interaction between neutrons situated
at height $z$ above the mirror surface and the bulk of the mirror
is given by
\begin{eqnarray}
V_a(z)= \int_{\mbox{\small mirror}} V(x',y',z+z') d^3
r'.\label{potgen}
\end{eqnarray}
The volume integral is calculated over the mirror bulk: $-\infty <
x',y' < \infty$, $z' < 0$ (in fact, over the neutron's vicinity
with the size of the order of a few $\lambda$ due to the
exponential convergence of these integrals). This integral can be
calculated explicitly. Thus a neutron with a given spin projection
to the vertical axis (orthogonal to the mirror surface) will be
affected by an additional exponential potential:
\begin{eqnarray}
V_a(z)=  \frac{g_p g_s}{4\pi} \frac{\pi \hbar \rho_m \lambda}{2m^2
c} \mbox{e}^{-z/\lambda}.\label{potfin}
\end{eqnarray}

Let us constrain the intensity of the axion-mediated interaction
$V_a$ from the experimental data. We will consider the potential
mentioned to be much weaker than the gravitational potential and
will therefore apply the perturbation theory. As previously
mentioned, the quantity extracted from the experimental data is a
classical turning point height $z_n$ for the neutron state $n$ in
the gravitational plus axion-mediated potential. In the absence of
any additional interaction a turning point $z_n$ is related to an
energy of the ``pure gravitational'' state $E_n$, as:
\begin{equation}\label{HE}
mgz_n=E_n
\end{equation}
In the presence of axion-mediated interaction  neutrons are
affected both by interaction with mirror (below) and with
scatterer (above). The height of the scatterer $\Delta h$ at which
the transmission through the slit starts for the $n$-th quantum
state is called $H_n$. Here, the resulting potential has the form:
\begin{equation}\label{withabs}
 W_a(z)=V_a(z)-V_a(H_n-z)
\end{equation}

In the presence of $W_a(z)$ the shift of a turning point height is
given by
\begin{equation}\label{withabsshift}
mg(z_n+\Delta z_n)+W_a(z_n+\Delta z_n)=E_n+\Delta E_n.
\end{equation}
In the first order of the perturbation theory expansion we get:
\begin{equation}\label{DeltaHVa}
\Delta z_n\approx\frac{\langle
\psi_n|W_a|\psi_n\rangle-W_a(z_n)}{mg+dW_a/dz(z_n)}.
\end{equation}

Here $\psi_n$ is a normalized wave-function of a neutron
gravitational state $n$. The analytical expressions for a
gravitational state wave-functions $\psi_n$ and gravitational
eigen-energies are well-known (see \cite{PRD} and references
therein):
\begin{eqnarray}\label{Ai}
 \psi_n(z)&=&\mathop{\rm Ai}((z-z_n)/l_0)/|\mathop{\rm
 Ai'}(-z_n/l_0)|\\
 \mathop{\rm
 Ai}(-z_n/l_0)&=&0.
\end{eqnarray}
Here $\mathop{\rm Ai}(z)$ is the Airy function \cite{Abramowitz}.

\begin{figure}
  \centering
 \includegraphics[width=115mm]{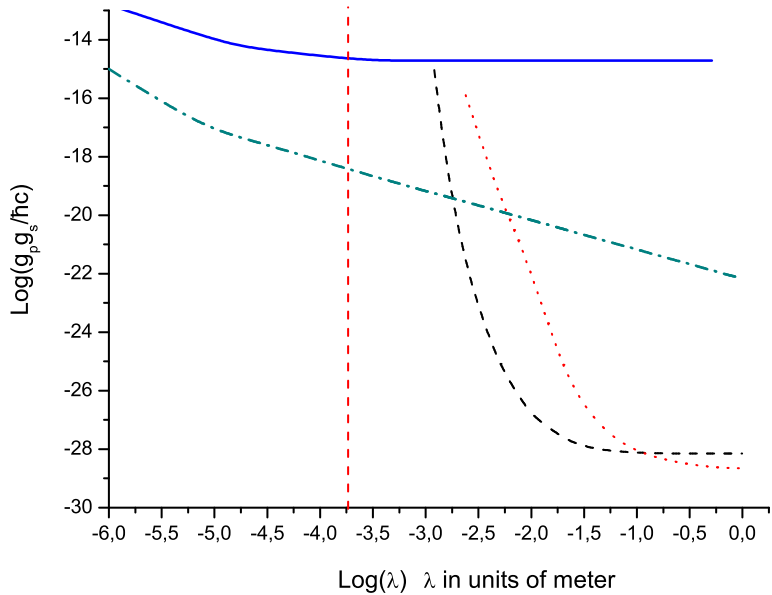}
 \caption{Constraints for the axion coupling. The solid line
indicates the present result, the dotted line corresponds to the
result of \cite{Yuod}, the dashed line shows the results of
\cite{Ritter, Ni} (approximately the same within graphical
accuracy) and the dash-dotted line illustrates the sensitivity
estimation for a future UCN experiment. The vertical dashed line
shows the characteristic range $\lambda$, for which a PVLAS signal
has been claimed.}\label{FigLimits}
\end{figure}

Expression (\ref{DeltaHVa}) allows the axion-mediated interaction
intensity to be constrained using the neutron flux measurement,
passed between the mirror and the scatterer in the presence of the
Earth's gravitational field \cite{EPJC}. The experimental values
of $z_n$ were established within 25\% accuracy for the first two
states; this can be considered as the upper limit for $\Delta
z_n$. The constraints for $g_s g_p/(\hbar c)$, obtained with the
above-mentioned data, are shown in Figure~\ref{FigLimits}. In the
same plot we show the existing experimental constraints from
refs.~\cite{Yuod, Ni, Ritter}. Note that in these experiments the
polarized particles were electrons whereas in our experiments the
polarized particle is neutron.

The signal reported by the PVLAS collaboration  corresponds to the
axion mass $1<M_A<1.5$ meV ($130< \lambda < 200$ $\mu$~m). In this
domain $\lambda/l_0\sim 30\gg 1$, so the expression
(\ref{DeltaHVa}) can be written as:
\begin{equation}\label{DeltaHlarge}
\Delta z_n\approx \frac{g_p g_s}{3\pi mg} \frac{\pi \hbar \rho_m
z_n}{2m^2 c}
\end{equation}
It follows from our results that $g_s g_p/(\hbar c)<2 \cdot
10^{-15}$ in this range of axion masses.


We should mention that a further increase in sensitivity could be
achieved, either by increasing the statistics, by using highly
excited quantum states, or by using the more intense UCN sources
now being developed, or by increasing observation time $T$ (for
experiments with specular traps without scatterer during storage
of the UCN) \cite{NOVA,NIM2006}. In the future experiment we plan
to use polarized neutrons and we expect to improve the relative
accuracy of the energy measurements for the quantum states by, at
least, two orders of magnitudes (dash-dotted line in
figure~\ref{FigLimits}). In this case possible false effects
caused by small magnetic impurities in
surface have to be carefully investigated.\\


We have thus established a constraint for the axion
monopole-dipole spin-matter coupling $g_s g_p/(\hbar c)$ in the
axion mass range of $0.1<M_A<200$ meV from the measurement of the
spatial distribution of UCN, passing along a horizontal mirror in
the presence of the Earth's gravitational field. In the axion mass
domain of $1<M_A<1.5$ meV, where the positive signal of the PVLAS
experiment was reported, we found that $g_s g_p/(\hbar c)<2 \cdot
10^{-15}$. The range of the axion masses studied was previously
out of reach of experimental study in the domain of spin-matter
coupling.

This limit can be improved by a few orders of magnitude in future
experiments with polarized UCN trapped inside the
gravitational states.\\

We would like to thank Carlo Rizzo for useful comments on the
PVLAS data, Stavros Katsanevas for stimulating discussions,
Hartmut Abele, Alexander Westphal and other members of the GRANIT
collaboration for their help.

We are grateful to the French Agence Nationale de la Recherche
(ANR) for supporting this project.


\begin{references}


\bibitem{AxTheo1} R. Peccei, H. Quinn, Phys. Rev. Lett. \textbf{38} 1440, (1977).
\bibitem{AxTheo2} S. Weinberg, Phys. Rev. Lett. \textbf{40}, 223 (1978).
\bibitem{AxTheo3} F. Wilczek, Phys. Rev. Lett. \textbf{40}, 279 (1978).
\bibitem{AxTheoRev} S.-L. Cheng et al., Phys. Rev. \textbf{D 52}, 3132
(1995).
\bibitem{Rosenberg} L.J. Rosenberg, K.A. van Bibber, Phys. Rep.
\textbf{325}, 1 (2000).
\bibitem{PDG} H. Murayama, G. Raffelt, C. Hagmann, K. van Bibber, L.J.
Rosenberg, Phys. Lett. \textbf{B 592}, 389 (2004).
\bibitem{PVLAS} E. Zavattini, G. Zavattini, G. Ruoso, E. Polacco, E. Milotti,
M. Karuza, U. Gastaldi, G. Di Domenico, F. Della Valle, R. Cimino,
S. Carusotto, G. Cantatore, and M. Bregant, Phys. Rev. Lett.
\textbf{96}, 110406 (2006).
\bibitem{NatureLam} S. Lamoreaux, Nature \textbf{441}, 31 (2006).
\bibitem{CAST} K. Zioutas et al. [CAST Collaboration], Phys. Rev. Lett.
\textbf{94}, 121301 (2005).
\bibitem{Anticontr} E. Masso and J. Redondo, JCAP \textbf{0509},
015 (2005) ; E. Masso and J. Redondo, arXiv:hep-ph/0606163,
to appear in Phys. Rev. Lett.;\\
P. Jain and S. Mandal, arXiv:astro-ph/0512155; \\
J. Jaekel, E. Masso, J. Redondo, A. Rigwal, and F. Takashani,
arXiv:hep-ph/0605313.
\bibitem{Antoniadis} I. Antoniadis, A. Boyarsky, O. Ruchayskiy,
arXiv:hep-ph/0606306.
\bibitem{Moody} J.E. Moody and F. Wilczek, Phys. Rev. \textbf{D30}, 130 (1984).
\bibitem{Yuod} A.N. Youdin, D. Krause, Jr., K. Jagannathan, L.R. Hunter,
S.K. Lamoreaux, Phys. Rev. Lett. \textbf{77}, 2170 (1996).
\bibitem{Ni} Wei-Tou Ni, Sheau-shi Pan, Hsien-Chi Yeh, Li-Shing Hou,
 Juling Wan, Phys. Rev. Lett. \textbf{82}, 2439 (1999).
\bibitem{Ritter} R.C. Ritter, L.I. Winkler, and G.T. Gillies, Phys. Rev. Lett.
\textbf{70}, 701 (1993).
\bibitem{Nature1} V.V. Nesvizhevsky, H.G. B\"orner, A.K. Petoukhov, H. Abele,
S. Bae{\ss}ler, F.J. Rue{\ss}, Th. St\"oferle, A. Westphal, A.M.
Gagarski, G.A. Petrov, A.V. Strelkov, Nature \textbf{415}, 297
(2002).
\bibitem{PRD1} V.V. Nesvizhevsky, H.G. B\"orner, A.M. Gagarski, A.K. Petoukhov,
G.A. Petrov, H. Abele, S. Bae{\ss}ler, G. Divkovic, F.J. Rue{\ss},
Th. St\"oferle, A. Westphal, A.V. Strelkov, K.V. Protasov, A.Yu.
Voronin,, Phys. Rev. \textbf{D67} 102002 (2003).
\bibitem{EPJC} V.V. Nesvizhevsky, A.K. Petoukhov, H.G. B\"orner, T.A. Baranova,
A.M. Gagarski, G.A. Petrov,  K.V. Protasov, A.Yu. Voronin, S.
Bae{\ss}ler, H. Abele, A. Westphal, L. Lucovac, Eur. Phys. J.
\textbf{C40}, 479 (2005).
\bibitem{NIM} V.V. Nesvizhevsky, H.G. B\"orner, A.M. Gagarski, G.A. Petrov,
A.K. Petoukhov, H. Abele, S. Bae{\ss}ler, Th. St\"oferle, S.M.
Soloviev, Nucl. Instr. Meth. in Phys. Res. \textbf{A440}, 754
(2000).
\bibitem{PANIC} H. Abele, S.Bae{\ss}ler, H.G. B\"orner, A.M. Gagarski,
V.V. Nesvizhevsky, A.K. Petoukhov, K.V. Protasov, A.Yu. Voronin,
A. Westphal, ``Gravitationally bound quantum states of neutrons:
applications and perspectives'', PANIC-2005, Santa-Fe, NM, USA,
October 24--28, 2005.
\bibitem{NOVA} V.V. Nesvizhevsky and K.V. Protasov, ``Quantum states of
neutrons in the Earth's gravitational field: state of the art,
applications, perspectives'', In {\em Trends in Quantum Gravity
Research}, ed. David C. Moore, (Nova Science Publishers Inc., New
York, 2006) p. 65.
\bibitem{Abramowitz} M. Abramowitz and  I.E. Stegun {\it Handbook of mathematical
Functions} (Dover Publ., New York, 1965).
\bibitem{PRD} A.Yu. Voronin, H. Abele, S. Bae{\ss}ler, V.V. Nesvizhevsky, A.K.
Petukhov , K.V. Protasov, A. Westphal, Phys. Rev. \textbf{D73},
044029 (2006).
\bibitem{NIM2006} V.V. Nesvizhevsky, Nucl. Instr. Meth. in Phys. Res.
\textbf{A557}, 576 (2006).

\end{references}
\end{document}